
\magnification 1200
%

%
\font\eightrm=cmr8
\font\eighti=cmmi8
\font\eightsy=cmsy8
\font\eightbf=cmbx8
\font\eighttt=cmtt8
\font\eightit=cmti8
\font\eightsl=cmsl8
\font\sixrm=cmr6
\font\sixi=cmmi6
\font\sixsy=cmsy6
\font\sixbf=cmbx6
\catcode`@11
\newskip\ttglue

\def\eightpoint{\def\rm{\fam0\eightrm}
\textfont0=\eightrm \scriptfont0=\sixrm \scriptscriptfont0=\fiverm
\textfont1=\eighti \scriptfont1=\sixi \scriptscriptfont1=\fivei
\textfont2=\eightsy \scriptfont2=\sixsy \scriptscriptfont2=\fivesy
\textfont3=\tenex \scriptfont3=\tenex \scriptscriptfont3=\tenex
\textfont\itfam=\eightit \def\it{\fam\itfam\eightit}
\textfont\slfam=\eightsl \def\sl{\fam\slfam\eightsl}
\textfont\ttfam=\eighttt \def\tt{\fam\ttfam\eighttt}
\textfont\bffam=\eightbf
\scriptfont\bffam=\sixbf
\scriptscriptfont\bffam=\fivebf \def\bf{\fam\bffam\eightbf}
\tt \ttglue=.5em plus.25em minus.15em
\normalbaselineskip=6pt
\setbox\strutbox=\hbox{\vrule height7pt width0pt depth2pt}
\let\sc=\sixrm \let\big=\eightbig \normalbaselines\rm}
\newinsert\footins
\def\newfoot#1{\let\@sf\empty
  \ifhmode\edef\@sf{\spacefactor\the\spacefactor}\fi
  #1\@sf\vfootnote{#1}}
\def\vfootnote#1{\insert\footins\bgroup\eightpoint
  \interlinepenalty\interfootnotelinepenalty
  \splittopskip\ht\strutbox 
  \splitmaxdepth\dp\strutbox \floatingpenalty\@MM
  \leftskip\z@skip \rightskip\z@skip
  \textindent{#1}\footstrut\futurelet\next\fo@t}
\def\fo@t{\ifcat\bgroup\noexpand\next \let\next\f@@t
  \else\let\next\f@t\fi \next}
\def\f@@t{\bgroup\aftergroup\@foot\let\next}
\def\f@t#1{#1\@foot}
\def\@foot{\strut\egroup}
\def\footstrut{\vbox to\splittopskip{}}
\skip\footins=\bigskipamount 
\count\footins=1000 
\dimen\footins=8in 

\def\ref#1{$^{#1}$}
\def\flex{\raise 6pt\hbox{$\leftrightarrow $}\! \! \! \! \! \! }
\def\oversome#1{ \raise 8pt\hbox{$\scriptscriptstyle #1$}\! \! \! \! \! \! }
\def\tr{ \mathop{\rm tr}}

\newbox\bigstrutbox
\setbox\bigstrutbox=\hbox{\vrule height10pt depth5pt width0pt}
\def\bigstrut{\relax\ifmmode\copy\bigstrutbox\else\unhcopy\bigstrutbox\fi}
\def\refer[#1/#2]{ \item{#1} {{#2}} }
\def\rev<#1/#2/#3/#4>{{\it #1\/} {\bf#2}, {#3}({#4})}
\def\boxit#1{\vbox{\hrule\hbox{\vrule\kern3pt
\vbox{\kern3pt#1\kern3pt}\kern3pt\vrule}\hrule}}

\def\2figure#1#2#3#4{\vbox{ \hrule width#1truecm \hbox{\vrule height#2truecm
\hskip #1truecm
\vrule height#2truecm }\hrule width#1truecm \hbox{\vrule\vbox{\hsize #1truecm
\baselineskip=10pt
\noindent\strut#3}\vrule}\hrule width#1truecm
\hbox{\vrule\vbox{\hsize #1truecm
\baselineskip=10pt
\noindent\strut#4}\vrule}\hrule width#1truecm  }}
\def\3figure#1#2#3#4#5{\vbox{ \hrule width#1truecm \hbox{\vrule height#2truecm
\hskip #1truecm
\vrule height#2truecm }\hrule width#1truecm \hbox{\vrule\vbox{\hsize #1truecm
\baselineskip=10pt
\noindent\strut#3}\vrule}\hrule width#1truecm
 \hbox{\vrule\vbox{\hsize #1truecm
\baselineskip=10pt
\noindent\strut#4}\vrule}
\hrule width#1truecm \hbox{\vrule\vbox{\hsize #1truecm
\baselineskip=10pt
\noindent\strut#5}\vrule}\hrule width#1truecm  }}

\def\sqr#1#2{{\vcenter{\hrule height.#2pt
   \hbox{\vrule width.#2pt height#1pt \kern#1pt
    \vrule width.#2pt}
    \hrule height.#2pt}}}


\def\smin{\,\raise 0.06em \hbox{${\scriptstyle \in}$}\,}
\def\smsubset{\,\raise 0.06em \hbox{${\scriptstyle \subset}$}\,}

\def\Natural{\hbox{\hskip 1.5pt\hbox to 0pt{\hskip -2pt I\hss}N}}

\def\Rational{\hbox{\hbox to 0pt{\hskip 2.7pt \vrule height 6.5pt
                                  depth -0.2pt width 0.8pt \hss}Q}}
\def\Real{\hbox{\hskip 1.5pt\hbox to 0pt{\hskip -2pt I\hss}R}}
\def\Complex{\hbox{\hbox to 0pt{\hskip 2.7pt \vrule height 6.5pt
                                  depth -0.2pt width 0.8pt \hss}C}}

\def \E {{{\rm e}}}


\def \1ok{{1\over \kappa ^2} }

\def \3dslim {{\rm DS}\!\!\!\!\!\!\!\!\lim }
\def \4dslim {{\rm DS}\!\!\!\!\!\!\!\!\!\!\lim }
\def \tr {{\rm tr}\, }

\def \2kk{\left( \matrix {2k\cr k\cr }\right) }
\def \Rs4{{R^k\over 4^k} }

\def \1ok{{1\over \kappa ^2} }



\hsize 6truein
\vsize 8.5truein
\baselineskip 13.3 truept
\def \E {{{\rm e}}}

\def \0j {j_{{}_0} }
\def \1j {j_{{}_1} }
\nopagenumbers
\centerline{\bf TOWARDS A SOLUTION OF TWO-DIMENSIONAL QCD}

\vskip .45truecm
\centerline{ E. ABDALLA}
\vskip .45truecm
\centerline { International Centre for Theoretical Physics - ICTP}
\centerline {34100 Trieste, Italy }
\vskip .45truecm

\centerline {\bf ABSTRACT}
\vskip -.3cm
$$\vbox{\baselineskip 10 truept  \eightpoint
{\hsize 4.8truein {\vbox { Using the well-known result for the fermionic
determinant in terms of a WZW theory, we write QCD$_2$ in bosonized form.
After some manipulations we give two versions of the theory, where it is
factorized as a product of the conformally invariant WZW models, ghost terms,
and a WZW action perturbed off criticality. We first prove that the latter
is an integrable model. Furthermore, as a consequence of the BRST analysis,
there are several constraints. Second-class constraints also show up.
Using the integrability condition we find higher conservation laws,
their action on the asymptotic states, and we propose an exact $S$-matrix for
the physical excitations. The vacuum structure is analysed, and we
prove that a finite number of different vacua exist. An outline of possible
generalizations of the procedure is given.}}}}$$
\vskip .2truecm

The QCD$_2$ Lagrangian with massless quarks and the corresponding partition
function are given by the expressions
$$\eqalignno{
{\cal L} &= -{1\over 4}\tr F_{\mu\nu}F^{\mu\nu} + \overline \psi
(i \!\not \!
\partial + e \!\not \!\! A)\psi \quad ,&(1)\cr
{\cal Z} &= \int {\cal D}\psi
{\cal D}\overline \psi {\cal D}A_\mu \, \E ^{i\int {\rm d}^2 z\,{\cal L} }
\quad , &(2)\cr}
$$
with the notation defined in [1].
The bosonized version of the theory was obtained by rewriting the fermionic
determinant $\det i\!\not\!\! D$ as a bosonic functional integral as
$${\det i\not\!\!D\over\det i\not\!\partial}
\equiv\E^{iW[A]}=\int{\cal D}g\,\E^{iS_F[A,g]}\quad ,\eqno(3)
$$
where $W[A]\equiv -\Gamma[UV]$, with $\Gamma$ the WZW action, and the
effective action $S_F[A,g]$ is the gauged WZW action. The external sources
will be dropped in the following (see [1] and [2] for details).
The non-linearity in the gauge-field interaction can be disentangled by
means of the identity
$$
\E^{-{i\over 4}\int{\rm d}^2z\,\tr F_{\mu\nu}F^{\mu\nu}}=\int{\cal D}E\,
\E^{-i\int{\rm d}^2z\,\left[{1\over 2}\tr E^2+{1\over 2}\tr EF_{+-}\right]}
\quad,\eqno(4)
$$
where $E$ is a matrix-valued field. Thus we arrive, for the partition
function, at the expression
$$ {\cal Z}= \int {\cal D}E{\cal D}
U{\cal D}V{\cal D}g \times \E^{i\Gamma[UgV] -i(c_V+1) \Gamma[UV] -i \int
{\rm d}^2 z \,\tr
[{1\over 2} E^2 + {1\over 2}E F_{+-}] }\quad ,\eqno(5)
$$
where $c_V$ is the quadractic Casimir. We should also include a term $\, m\,\tr
(g\!+\!g^{^{-1}})$ in the effective action, if we were considering massive
fermions,
but we shall avoid such a complication and consider only the massless case for
the moment.

Gauge fixing is another ingredient and, in fact, the process of introducing
ghosts here is standard. We perform the procedure implicitly, until it is
necessary to explicitly take into account the ghost degrees of freedom. Up to
considerations concerning the spectrum, our manipulations do not explicitly
depend on the gauge fixing/ghost system, and we keep it at the back of our
minds and formulae.

It is not difficult to see that the field $\widetilde g$ decouples (up to the
BRST condition) after defining a new gauge-invariant field $\widetilde g=UgV$.
Using the invariance of the Haar measure, ${\cal D}g={\cal D}\widetilde g$,
the partition function turns into
$${\cal Z}=\int{\cal D}\widetilde g\,\E^{i
\Gamma[\tilde g]}{\cal D}E{\cal D}U{\cal D}V{\cal D}({\rm ghosts})
\E^{-i(c_V+1)\Gamma[UV]-i\int{\rm d}^2z\,\tr[{1\over 2}E^2+{1\over 2}E
F_{+-}]+iS_{\rm ghosts} }\, ,\eqno(6)
$$
where the $A_\mu$ variables are given in terms of the two-matrix-valued fields
$U$ and $V$, as usual.\ref 1

In the way the gauge-field strength $F_{+-}$ is presented, it hinders
further developments; however if we write it in terms of the $U$ and $V$
potentials, we arrive at the identity
$
\tr EF_{+-}={i\over e}\tr UEU^{-1}\partial_+(\Sigma\partial_-\Sigma^{-1})$.
We have used the variable $\Sigma=UV$; this will imply a further
factorization of the partition function. In fact, $\Sigma$ is a more natural
candidate for representing the physical degrees of freedom, since $U$ and $V$
are not separately gauge-invariant. We redefine $E$, taking once more
advantage of the invariance of the Haar measure, in such a way that the
effective action depends only on the combination $\Sigma$. The variables
$U$ and $V$ will then appear separately only in the source terms.
If we eventually choose the light-cone gauge, we will have  $U=1$ and
$A_+=0$. It is natural to redefine variables as $\widetilde E'=UEU^{-1},{\cal
D} E={\cal D}\widetilde E'$. Notice that already at this point the $E$
redefinition implies, in terms of the gauge potential, an infinite gauge tail,
which captures the possible gauge transformations. It is also convenient to
make the rescaling $\widetilde E'=2e(c_V+1)\widetilde E$, with a constant
Jacobian. In terms of the field $\widetilde E$, consider the change of
variables
$$
\partial_+\widetilde E = {i\over 4\pi} \beta^{-1}\partial_+\beta \quad ,\quad
{\cal D}\widetilde E = \E^{-ic_V\Gamma[\beta]}{\cal D} \beta\quad ,\eqno(7)
$$
introducing $\beta$, the analogous of a loop variable. Now we use the
identity
$$\Gamma[UV]=\Gamma[U]+\Gamma[V]+
{1\over 4\pi}\tr\,\int{\rm d}^2x\, U^{-1}\partial_+UV
\partial_- V^{-1}\eqno(8)$$
to transform the $\beta \Sigma$ interaction into terms that
can be handled in a more appropriate fashion.
After replacement of $E$  in terms of $\beta$ in the partition function, and
using (8) for $\Gamma [\beta \Sigma]$, we arrive at
$$\displaylines{ {\cal Z}=\int{\cal D}\widetilde g {\cal D}U{\cal D}
({\rm ghosts}) {\cal D} \widetilde\Sigma
\,\E^{i\Gamma[\tilde g]-i(c_V+1)\Gamma[\widetilde\Sigma] +iS_{{\rm ghosts}}}
\times \hfill\cr
\hfill \int {\cal D} \beta\, \E^{i\Gamma[\beta] + {\mu^2i\over 2}\tr \int
{\rm d}^2 z\,[\partial_+^{-1}(\beta^{-1}\partial_+\beta)]^2}
\, ,\quad (9)\cr}
$$
where we defined the massive parameter $\mu = (c_V+1)e/2\pi$ and the field
$\widetilde \Sigma = \beta\Sigma$, and we used the
Haar invariance of the $\Sigma$ measure.

Up to BRST constraints and  source terms, the above generating functional
factorizes in terms of a conformally invariante theory for $\widetilde
g$, which
represents a gauge-invariant bound state of the fermions, of a second
conformal field  theory for $\widetilde \Sigma$, which represents some gauge
condensate, and of an off-critically perturbed conformal field theory for the
$\beta$ field, which also describes a gauge-field condensate, which we
interpret as an analogue of the Wilson-loop variable in view of the change of
variables (7).  The conformal field theory representing $\widetilde\Sigma$
displays an action with a negative sign. Therefore we have to carefully
take into account the BRST constraints in order to arrive at a positive metric
Hilbert space. This is reminiscent of the commonly encountered negative metric
states of gauge theories, and appeared already in the Schwinger model.\ref{7}
In that case the requirement that the longitudinal current containing the
negative metric field vanishes, implies the decoupling of the unwanted fields
from the physical spectrum. The only trace of such massless fields is the
degeneracy of the vacuum. In that case, the chiral densities commute with the
longitudinal part of the current, and it is possible to build operators
carrying non-vanishing fermion number and chirality. They are, however,
constant operators commuting with the Hamiltonian, and the ground state turns
out to be infinitely degenerate. There are definite vacua superpositions where
the above  states are just phases -- the so-called $\theta$-vacua. Here we
will learn that there is a finite number of such vacua.

\vskip .7truecm
\noindent {\it Integrability and duality }
\vskip .3truecm
\nobreak
At the Lagrangian level we are left, for the massive part of
the theory, with the perturbed WZW action\ref{2,3}
$$
\eqalign{
S &= \Gamma[\beta] + {1\over 2} \mu^2 \tr \int {\rm d}^2z\, \left[\partial_+^
{-1}(\beta^{-1} \partial_+ \beta)\right] ^2\quad ,\cr
&= \Gamma[\beta] + {1\over 2} \mu^2 \Delta(\beta)\quad .\cr}\eqno(10)
$$

We will look for the Euler--Lagrange equations for $\beta$. It is not difficult
to find the variations:
$$
\eqalignno{
\delta \Gamma[\beta ]&= \left[{1\over 4\pi}\partial_-(\beta^{-1}\partial_+
\beta)\right]\beta^{-1}\delta\beta\quad ,&(11a)\cr
\delta \Delta(\beta) & = 2\Big( \partial_+^{-1} (\beta^{-1}\partial_+\beta)
 - \big[ \partial_+^{-2} (\beta^{-1}\partial_+\beta),(\beta^{-1}\partial_+
\beta) \big] \Big)\beta^{-1}\delta \beta\quad .&(11b)\cr}
$$

Collecting the terms, we find it useful to define the current components
$$
\eqalign{
J_+&= \beta^{-1}\partial_+ \beta\quad ,\cr
J_-&= -4\pi\mu^2\partial_+^{-2}J_+ =-4\pi\mu^{2}
\partial_+^{-2}(\beta^{-1}\partial_+\beta)\quad ,\cr}\eqno(12)
$$
which summarize the $\beta $ equation of motion as a zero-curvature condition
given by
$$
[{\cal L},{\cal L}]=[\partial_++J_+,\partial_- +J_-]=\partial_-
J_+-\partial_+J_-+[J_-, J_+ ] =0 \quad .\eqno(13)
$$
This is the integrability condition for the Lax pair\ref{7}
$$
{\cal L}_\mu M =0 \quad  , \quad {\rm with } \quad {\cal L}_\mu = \partial _
\mu - J_\mu \quad ,\eqno(14)
$$
where $J_\pm =J_0\pm J_1$ and $M$ is the monodromy matrix.
This is not a Lax pair as in the usual non-linear $\sigma$-models,
where $J_\mu$ is a conserved current, and where we obtain a conserved
non-local charge from (13), as well as higher local and non-local
conservation laws.\ref{7} However, to a certain extent, the
situation is simpler in the present case, due to the rather unusual form of
the currents (12), which permits us to write the commutator appearing in
(13) as a total derivative, in such a way that in terms of the current
$J_-$ we have
$$
\partial_+\left(4\pi\mu^2J_- +\partial_+\partial_-J_-+[J_-,
\partial_+J_-]\right) \equiv \partial _+ I_- = 0\quad . \eqno(15)
$$
Therefore the quantity $I_-$, as defined above,
does not depend on $x^+$, and it is a simple matter to derive an infinite
number of conservation laws from the above. These are higher conservation
laws.

This means that two-dimensional QCD is an integrable system!
Moreover, it corresponds to an off-critical perturbation of the WZW action. If
we write $\beta=\E^{i\phi}\sim 1+i\phi$,  we verify that the perturbing term
corresponds to a mass term for $\phi$. The next natural step is to obtain the
algebra obeyed by $I_-$, and its representation, in spite of the
difficulty presented by the non-locality of the perturbation.

Consider the $\Delta$-term of the action (10). We rewrite it, introducing the
integral over a Gaussian field $C_-$ as
$$
\E^{{i\over 2}\mu^2\Delta}=\int{\cal D}C_-\,\E^{i\int{\rm d}^2x\,{1\over 2}\tr
(\partial_+C_-)^2-\mu\tr\int{\rm d}^2x\,C_-(\beta^{-1}\partial_+\beta)}\quad,
\eqno(16)
$$
where the left-hand side is readily obtained by completing the square in the
right-hand side.

In this case we have the action (for a local counterpart and further
discussions see [2] and [3])
$$
S = \Gamma[\beta] + i\mu\tr \int {\rm d}^2 x\,C_- \beta ^{-1}\partial_+
\beta + {1\over 2} \tr \int {\rm d}^2x\, (\partial _+C_-)^2 \quad .\eqno(17)
$$

Now, the canonical quantization proceeds straightforwardly, and the relevant
phase-space  expressions are obtained for $J_-$ in (12); because of the
$C_-$ equation of motion, these read
$$
\eqalignno{
J_- &= -4\pi \mu^2 \partial _+^{-2} (\beta^{-1} \partial _+ \beta) =
4i\pi \mu C_-\quad ,&(18a)\cr
\Pi_- & = \partial _+ C_-\quad ,&(18b)\cr}
$$
while the $\beta$-momentum is given by
$$
\widetilde{\hat \Pi}_{ji}  = {1\over 4\pi} \partial _0 \beta^{-1}_{ji} +
i\mu(C_-\beta^{-1})_{ji}\quad ,\eqno(19)
$$
where the hat above $\Pi$ means that we have neglected the WZW
contribution as before, and as a consequence\ref 7
$$\left\{ \widetilde{\hat \Pi}_{ji}(t,x), \widetilde{\hat \Pi}_{lk}
(t,y)\right\} =-{1\over 4\pi}\left( \partial_1\beta^{{}^{-1}}_{_{jk}}
\beta^{{}^{-1}}_{_{li}}-\partial_1\beta^{{}^{-1}}_{_{li}}
\beta^{{}^{-1}}_{_{jk}}\right)\delta(x-y)\quad .\eqno(20)
$$

We now come to the point where we should consider the quantization of the
symmetry current $I_-$. Since the only massive scale is the coupling constant,
we have to  consider the weak coupling limit. In such a case, we need the
short-distance expansion of the
current $J_-=-4\pi\mu^2\partial_+^{-2}(\beta^{-1}\partial_+\beta)$ with
itself. Since the short-distance expansion is compatible with the weak
coupling limit, where the theory is conformally invariant, Wilson expansions
can be dealt with in the usual way, and no anomaly is found.\ref 2

Since $I_{ij}$ is a right-moving field operator, it is natural to assume,
in view of the Poisson algebra (see [1,2,3] for details), that it obeys an
algebra given by
$$
I_{ij}(x^-) I_{kl}(y^-) = (I_{kj}\delta_{il}-I_{il}\delta_{kj})(y^-)
{1\over x^--y^-}
- {c_V+1\over 2\pi}{\delta^{il}\delta^{kj}\over(x^--y^-)^2}\quad .\eqno(21)
$$

For $J_{-ij}$ we are forced into a milder assumption. Since $I_ {ij}$ is a
right-moving field operator, the equal-time requirements in commutators
involving it is  superfluous, and we get an operator-product algebra of the
type
$$
I_{-ij}(x^-)J_{-kl}(y^+,y^-)=(J_{-kj}\delta_{ij}-J_
{-il}\delta_{kj})(y^+,y^-){1\over x^--y^-}+2{\delta^{il}\delta^{kj}\over(x^-
-y^-)^2}\quad .\eqno(22)
$$

Some conclusions may be drawn for $J_-$. It is clear that $\partial
_+J_-$ cannot be zero; however, in view of (22), we conclude that left
$(+)$ derivatives of this current are primary fields, since
$$
I_{ij}\partial_+^nJ_{-kl}={1\over x^--y^-}\left(\partial_+^nJ
_{-il}\delta_{kj}-\partial_+^nJ_{-kl}\delta_{il}\right)\quad .\eqno(23)
$$

Therefore, we expect an affine Lie algebra for $I(x^+)$, and $\partial_+
^n J_-$ should be primary fields depending on parameters $x^-$.

Such an underlying structure is a rather unexpected result, since
it arose out of a non-linear relation obeyed by the current, which
can be traced back to an integrability condition of the model. Moreover, the
theory has an explicit mass term -- although free massive fermionic theories as
well as some off-critical perturbations of conformally invariant theories in
two dimensions may contain affine Lie symmetry algebras.

Consider the effective action
$$
S_{\rm eff} = \Gamma[\widetilde g] - (c_V+1)\Gamma[\Sigma]  +S_{\rm gh}+
\Gamma[\beta] -
{1\over 2}\mu^2 \int {\rm d}^2 x\, [\partial_+^{-1}(\beta^{-1}
\partial _+\beta)]^2  \, .\eqno(24)
$$
Let us start by first coupling the fields  $(\widetilde g, \Sigma,
{\rm ghosts})$ to external gauge fields
$
A^{^{\rm ext}}_-={i\over e}V_{_{\rm ext}}\partial_-V_{_{\rm ext}}^{-1}$ and
$A^{^{\rm ext}}_+={i\over e}U_{_{\rm ext}}^{-1}\partial_+U_{_{\rm ext}}$. Such
a coupling may be obtained by substituting each WZW functional in terms of a
gauged WZW theory\ref 7. In the case of ghosts a chiral rotation must be
performed.  Therefore, after such a  procedure and using again the invariance
of the Haar measure to substitute $U_ {_{\rm ext}}[\widetilde g,\Sigma]
V_{_{\rm ext}} \to [\widetilde g,\Sigma]$, one finds
that the partition function does not depend on the external gauge fields
just introduced. This signals the presence of constraints.\ref{4}
Similar results follow from arguments based on the BRST analysis of the
theory.\ref{5,6}
Thus, functionally differentiating the partition function once with
respect to $A^{^{\rm ext}}_+$ and separately with respect to
$A^{^{\rm ext}}_-$, and putting $A^{^{\rm ext}}_\pm =0$ we find the
constraints
$$
\eqalignno{
i\widetilde g\partial_-\widetilde g^{-1}-i(c_V+1)\Sigma\partial_-\Sigma^{-1}
+J_-({\rm ghosts})& \sim 0\quad ,&(25a)\cr
i\widetilde g^{-1}\partial_+\widetilde g-i(c_V+1)\Sigma^{-1}\partial_+\Sigma
+J_+({\rm ghosts})& \sim 0\quad ,&(25b)\cr}
$$
leading to two BRST charges $Q^{(\pm)}$ as discussed in refs. [4,5,6], which
are nilpotent. Therefore we find two first-class constraints.

The field $A^{^{\rm ext}}_+$ can also be coupled to the field $\beta$ instead
of $\widetilde g$, since the system $(\beta,\Sigma,{\rm ghosts})$ has
vanishing central charge too. In such a case we have to disentangle the
non-local interaction considering instead of the $\beta$-terms in (24), the
$\beta$ action
$$
S(\beta) = \Gamma[\beta]+ \int {\rm d}^2 x\, {1\over 2}(\partial _+C_-)^2 +
i\int {\rm d}^2 x \,\mu\, C_-\beta^{-1} \partial _+\beta\quad .\eqno(26)
$$

We make the minimal  substitution $\partial _+ \to\partial _+-ieA^{^{\rm ext}}
_+$, repeating the previous arguments for the $(\beta, \Sigma, {\rm ghosts})$
system, and we now arrive at the constraint (the minus gauging is not an
invariance if one includes the $\beta$ system):
$$
\beta \partial _-\beta ^{-1}+4i\pi\mu\beta C_-\beta^{-1} - i(c_V+1)\Sigma
\partial _-\Sigma^{-1} + J_-({\rm ghost}) \sim 0\quad .\eqno(27)
$$

One could na\"{\i}vely expect that, by repeating the previous arguments, one
obtains a system with a new set of first-class constraints. But if we
consider instead the equivalent system of constraints defined by the first set
(25$a$), together with the difference of the $(-)$ currents, i.e. (25$b$) and
(26) as given by
$$
\Omega_{ij}=(\beta \partial _-\beta^{-1})_{ij} + 4i\pi \mu (\beta C_-\beta
^{-1})_{ij} - (\widetilde g \partial _-\widetilde g^{-1})_{ij}\quad ,\eqno(28)
$$
one readily verifies that the latter cannot lead to a nilpotent
BRST charge due to the absence of ghosts. Therefore, it must be treated as a
second-class constraint. The Poisson algebra obeyed by $\Omega_{ij}$ is
$$
\eqalignno{
\left\{ \Omega_{ij}(t,x), \Omega_{kl}(t,y)\right\}&=(\widetilde \Omega_{il}
\delta_{kj} - \widetilde \Omega_{kj}\delta _{il})(t,x) \delta(x-y) + 2 \delta_
{il}\delta_{kj} \delta'(x-y)\, ,&(29a)\cr
{\widetilde \Omega} & = \widetilde g \partial _- \widetilde g^{-1}+
\beta\partial_-\beta^{-1}+4i\pi\mu\beta C_-\beta^{-1}\quad .&(29b)\cr}
$$
(Notice the change of sign in $\widetilde \Omega$.) Using the above, we can
thus define the undetermined velocities, and no further constraint is
generated.

The fact that the theory possesses second-class constraints is very
annoying, since these cannot be realized by the usual cohomology construction.
Therefore, instead of building a convenient Hilbert space, one has to modify
the dynamics, since the usual relation between Poisson brackets and commutators
is replaced by the relation between Dirac brackets and commutators. We will not
discuss this issue here (see [2,5,6])

In order to understand the nature of the conservation laws we consider
the conserved charges
$$Q^{(n)}=\oint dx^- [x^-]^n I_-\quad ,\eqno(30)$$
from which one obtains, for the asymptotic charge, as a consequence
of its Lorentz transformation properties, the expression
$$
Q^{(n)} \simeq (p_-)^n J \quad , \eqno(31)$$
where $J$ is the generator of the right transformations for the $\beta$ fields.
This would mean that the first charge, $Q^{(1)}$, is the generator of
right-$SU(N)$ transformations for the $\beta$ fields! In the quantum theory
there is no contribution from the short-distance expansion of $J_-$ and
$\partial_\mu J_-$, since divergences are too mild. The $SU(N)$ transformation
generators are simple to compute. One has, for
left-$SU(N)$ transformations:
$$\eqalignno{
J^{L_{SU(N)}}_+ &=  0\quad ,&\cr
J^{L_{SU(N)}}_- &=  -{1\over 4\pi}\partial_-\beta\beta^{-1}+i\mu\beta
C_-\beta^{-1}\quad ,&(32)\cr}$$
while for right-$SU(N)$ transformations we obtain
$$\eqalignno{
J^{ R_{SU(N)}}_- = & i\mu C_-+\left[ \partial_+C_-,C_-\right] \quad ,
\cr
J^{ R_{SU(N)}}_+ = & -{1\over 4\pi}\beta^{-1}\partial_+\beta
\quad . &(33)\cr}$$

For the first set, i.e. the left transformations, one finds that the currents
are equivalent to analogous chiral currents corresponding to free fields. On
the other hand, the right transformations lead to an infinite number of
conservation laws due to the presence of the Lax pair. See [8] for a
detailed discussion in terms of the dual theory, where the procedure
is even clearer.

The left indices, such as $a,b,c,d$ are thus free,
described by a trivial $S$-matrix, while the right indices, such as $i,j,k,l$
are described by an $SU(N)$ covariant integrable $S$-matrix, which has a
well-known classification.\ref{9} Indeed, $SU(N)$-invariant $S$-matrices of
the integrable type can be classified within five very definite types.\ref{9}
The first one is trivial. The third one is $O(N)$-invariant and does not
concern us here. The fourth and fifth types have a rather strange form, but
what decides for the $S$-matrix of the second type is the fact that in such
a case,  the particle/antiparticle backward scattering vanishes:
a characteristic of
scatterings where antiparticles are bound states of particles, or a
bound-state structure as in the $Z_N$ model appears (see chapter 8 of ref.
[7]).
In the present case, for large $N$, the interpretation of $\beta $ as a
loop variable implies $\beta^2\sim \beta$. We thus write the (unique) ansatz
for scattering of $\beta$ particles obeying the above requirements
$$\eqalignno{
\langle ai\theta_1, bj\theta_2\vert ck\theta_3dl\theta_4\rangle = \, &  \delta(
\theta_1-\theta_3)\delta(\theta_2-\theta_4)\delta_{ac}\delta_{bd}(\sigma_1(
\theta)\delta_{ik}\delta_{jl}+\sigma_2(\theta)\delta_{il}\delta_{jk}) +\cr
\, &\delta (\theta_1-\theta_4)\delta(\theta_2-\theta_3)\delta_{ad}\delta_{bc}
(\sigma_1(\theta)\delta_{il}\delta_{jk}+\sigma_2(\theta)\delta_{ik}\delta_{jl}
)\, ,&(34)\cr}$$
where $\theta=\theta_1-\theta_2$ is the relative rapidity.

The discussion of the theory is only complete after we discuss the properties
of the vacuum, described by the conformally invariant part of the action (24).
However, as pointed out in [6] it is described by a G/G topological field
theory, and the vacuum states, obeying the BRST conditions defined by the
constraints, are only two and are given in terms of the eigenvalue of the third
component of the $\widetilde g$ current. Building a level-1 Kac--Moody
algebra, there are, in the $SU(2)$-case, only two possible values for such an
eigenvalue, namely zero or one-half. In the general $SU(n)$ case, or for a
higher representation of the matter fields, a similar structure arises,
with a larger number of vacuum states.

The massive quark case is at present under study,\ref{10} and a rich constraint
structure arises in that case as well.

\vskip .75truecm
{\bf References}
\vskip.3truecm

\refer[[1]/E. Abdalla and M.C.B. Abdalla, hep-th/9503002, Phys. Rep., to
appear]

\refer[[2]/E. Abdalla and M.C.B. Abdalla, Int. J. Mod. Phys. {\bf A 10},
1611 (1995)]

\refer[[3]/E. Abdalla and M.C.B. Abdalla, Phys. Lett. {\bf B 337}, 347 (1994)]

\refer[[4]/D. Karabali and H.J. Schnitzer, Nucl. Phys. {\bf B 329}, 649 (1990)]

\refer[[5]/D. C. Cabra, K.D. Rothe, and F.A. Schaposnik,
Heidelberg preprint HD-THEP-95-31, hep-th/9507043]

\refer[[6]/E. Abdalla and K.D. Rothe, hep-th/9507114, to appear in Phys.
Lett.  B]

\refer[[7]/E. Abdalla, M.C.B. Abdalla and K.D. Rothe,
{\it Non-perturbative Methods in Two-Dimensional Quantum
Field Theory}, (World Scientific, Singapore), 1991]

\refer[[8]/E. Abdalla and M.C.B. Abdalla, CERN-TH/95-81, hep-th/9503235]

\refer[[9]/B. Berg, M. Karowski, V. Kurak and P. Weisz, Nucl. Phys.
{\bf B 134}, 125 (1978)]

\refer[[10]/E. Abdalla, M.C.B. Abdalla and K.D. Rothe, in preparation]

\end